\documentclass[onecolumn]{article}
\usepackage{epsfig,amssymb,amsfonts}

\begin{document}


\title{Photonic Crystals and Inhibition of Spontaneous Emission: An Introduction}




\author{D. G. Angelakis \\
{\it Centre for Quantum Computation, Department of Applied
Mathematics }\\
 {\it and Theoretical Physics, University of Cambridge,}\\
 {\it Wilberforce
Road, CB3 0WA, U.K} \and
P. L. Knight \\
{\it QOLS, Blackett Laboratory, Imperial College, London SW7 2AZ,
UK} \and
E. Paspalakis \\
{\it Materials Science Department, School of Natural Sciences}\\
 {\it University of Patras, Patras 265 04, Greece}
 }


\date{\today}
\maketitle

\begin{abstract}

In the first part of this introductory review we outline the
developments in photonic band gap materials from the physics of
photonic band gap formation to the fabrication and potential
applications of photonic crystals. We briefly describe the
analogies between electron and photon localization, present a
simple model of a band structure calculation and describe some of
the techniques used for fabricating photonic crystals. Also some
applications in the field of photonics and optical circuitry are
briefly presented. In the second part, we discuss the consequences
for the interaction between an atom and the light field when the
former is embedded in
 photonic crystals of a specific type, exhibiting a specific form
 of a gap in the density of states.
 We first briefly review the standard treatment (Weisskopf-Wigner theory)
  in describing the
 dynamics of spontaneous
 emission in free space from first principles, and then
 proceed by explaining the alterations needed to properly
 treat the case of a two-level atom embedded in
 a photonic band gap material.
\end{abstract}

\section{Introduction} \label{intro}

At the microscopic level, ordinary matter exhibits behaviour
analogous to light waves. When wave-like electrons scatter off
ions in crystalline materials, constructive interference between
different trajectories can cause electric currents to flow.
Conversely, disorder in such crystals can hinder electrical
conductivity, and for some energies the electrons become localized
in space, thus preventing their free flow in the form of
electrical currents . Although such ideas have been known since
the 1960s, it is only recently that physicists have begun to ask
whether similar effects can result in the localization of light in
a corresponding ``photonic crystal''.

The purpose of this article is twofold. In the first part we
outline the developments in photonic band gap (PBG) materials from
the physics of photonic band gap formation to the fabrication and
potential applications of photonic crystals. In the latter, we
specifically focus on the state of art 3-D structures exhibiting a
full band gap in their electromagnetic field density of states,
because as we believe, these are of great technological and
fundamental interest. In the second part, we discuss the
consequences for the interaction between an atom and the light
field when the former is embedded in dielectric materials, mostly
photonic crystals, which exhibit gaps in the density of states. We
first review the standard treatment to describe the dynamics of
spontaneous emission in free space and proceed by explaining the
alterations needed to properly illustrate the dynamics of
two-level systems embedded in a PBG material.

\section{Developments in photonic crystals}

\subsection{From electrons to photons}

\begin{figure}
\centerline{\psfig{figure=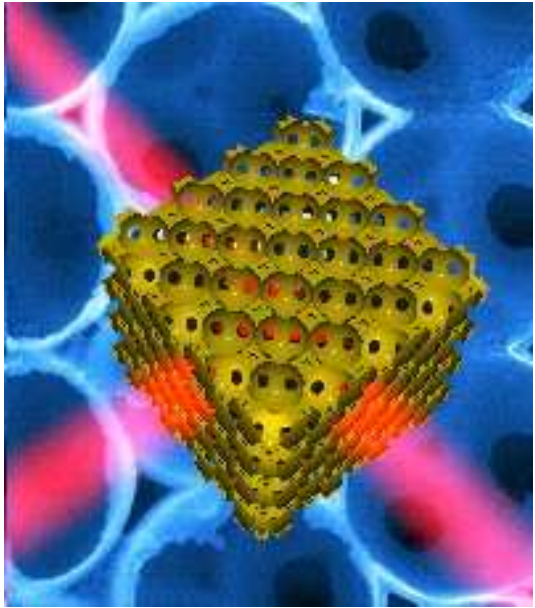,height=8.cm}} \caption { A
3-d silicon photonic crystal exhibiting a complete band gap for
incident light at 1.5 microns. It is constructed by growing
silicon inside the voids of a opal template of close-packed silica
spheres that are connected by the small ``necks'' formed during
sintering, followed by removal of the silica template-see Sec. 2
(courtesy of S. John's group, University of Toronto).}
\label{John_inv_opal}
\end{figure}

A normal crystal is a periodic array of atoms which scatters and
modifies the energy momentum relation of electrons, whereas a
photonic crystal is an ordered inhomogeneous medium characterized
by a spatially periodic dielectric constant, with lattice
parameter comparable to the wavelength of the light
\cite{Joannopoulosbook}. Fig. \ref{John_inv_opal} shows one such
microstructure in which a complete bandgap at 1.5 microns has been
observed.

\begin{figure}
\unitlength1cm
\begin{minipage}[t]{5cm}
\begin{picture}(5,5)
\psfig{figure=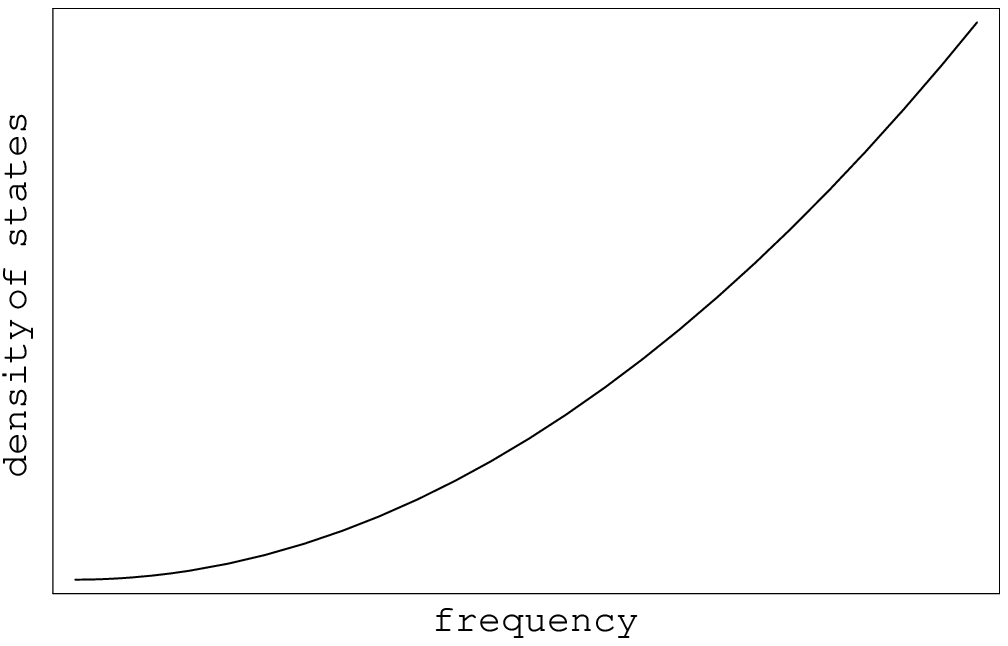,width=7.cm}
\end{picture}\par
\caption {The density of modes of the EM field in free space.}
\label{freespace_dos}
\end{minipage}
\hfill
\begin{minipage}[t]{6cm}
\begin{picture}(5,5)
\psfig{figure=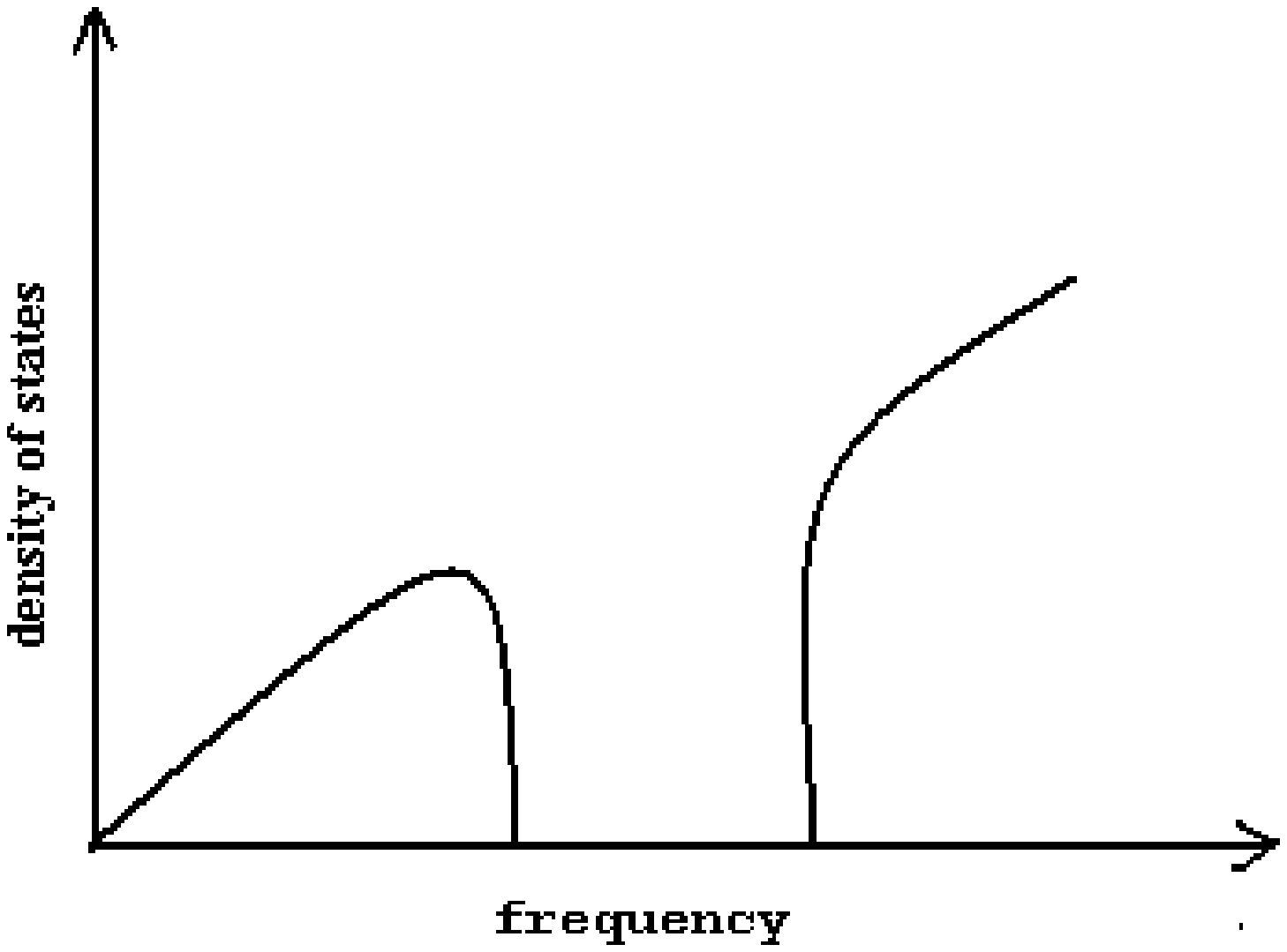,width=7.cm}
\end{picture}\par
\caption { The density of modes of the EM field in a PBG structure
with a broad gap.}\label{pbg_dos}
\end{minipage}
\end{figure}

Strictly speaking, such a structure has no allowed
electro-magnetic (EM) modes in the forbidden range of frequencies
or in other words the {\it density of states} of the propagating
photon modes is zero (see Figs. \ref{freespace_dos},
\ref{pbg_dos}). By contrast in free space (a cavity of infinite
volume) the density of modes varies as $\omega^{2}$ and exhibits
no gap. In a cavity of finite volume the density of states is
substantially modified for frequencies close to the cavity cut
off. Below the cut off the cavity sustains no modes at all, and
near above the cutoff, the density of states can be increased to
the continuum case.
The vanishing of the density of propagating photon modes within a
PBG means that, for the frequency range spanned  by the gap,
linear propagation of EM waves is {\bf forbidden in any direction}
in the PBG material. Thus, light incident on a PBG material with a
frequency from the gap region will be backscattered from the
material, independent of the angle of incidence. Strong
suppression of transmission with an associated peak in the
reflectivity at the characteristic frequencies is then an
experimental signature of a photonic band gap
\cite{Yablonovitch87a}.

To give a further insight on the physics of light localization we
will exploit the analogy between electron and light a bit further.
As is well known an electron in a disordered solid is described by
the following Schr\"{o}dinger equation:
\begin{eqnarray}
-\frac{\hbar^{2}}{2m^{\ast}} \nabla^2\psi({\bf x})+V({\bf
x})\psi({\bf x})=E\psi({\bf x}) \, , \label{intro1}
\end{eqnarray}
where $m^{\ast}$ is the electron's effective mass and $V({\bf x})$
is a potential that varies randomly in space. For sufficiently
negative energies $E$, the electrons may get trapped in regions
where the random potential is very deep. The rate at which
electrons tunnel out of the deep potentials depends on the
probability of finding nearby potential fluctuations into which
the trapped electron can tunnel. This rate increases as the
electron energy increases.

In the case of a monochromatic EM wave of frequency $\omega$
propagating in an inhomogeneous but non-dissipative dielectric
medium, the classical wave equation for the electric field is
\begin{eqnarray}
-\nabla^{2}{\bf E}+\frac{1}{c^2}\frac{{\partial{\bf
E}}^2}{{\partial t}^2}=\mu_{0} \frac{{\partial{\bf
P}}^2}{{\partial t}^2} \, ,
\end{eqnarray}
where ${\bf P}$ is the polarization of the medium. Assuming now
propagation in a linear medium, this leads to
\begin{eqnarray}
-\nabla^{2}{\bf \cal{E}}-\frac{\omega^2}{c^2}\epsilon_{fluct}({\bf
x}){\bf \cal{E}} =\epsilon_{D} \frac{\omega^2}{c^2}{\bf \cal{E}}
\, , \label{intro2}
\end{eqnarray}
where $\cal{E}$ is the slowing varying field amplitude. Also the
total dielectric constant has been separated into its average
value $\epsilon_{D}$ and a spatially fluctuating part
$\epsilon_{fluct}({\bf x})$. In a lossless material the dielectric
constant $\epsilon({\bf x})$ is everywhere real and positive
\footnote{All the dielectric materials described in this article
will assumed to be completely lossless.} and plays a role
analogous to the random potential $V({\bf x})$ in the
Schr\"{o}dinger equation. It scatters the EM wave.

Comparing now the above two equations (\ref{intro1}) and
(\ref{intro2}) we observe the following differences: Firstly, the
quantity $\epsilon_{D} \frac{\omega^2}{c^2}$ which plays the role
of an energy eigenvalue, is always positive, which precludes the
possibility of elementary bound states of
light in deep negative potential wells.\\
Secondly the mode frequency $\omega$ multiplies the scattering
potential $\epsilon_{fluct}({\bf x})$ and in contrast to an
electronic system, where localization is increased by lowering the
electron energy, lowering the photon energy leads to a complete
disappearance of scattering. In addition, looking at the opposite
high-frequency limit, geometric ray optics becomes valid and
interference corrections to optical transport becomes less and
less effective. These simply mean that in both cases the normal
modes of the EM field are extended, not localized. Finally, the
condition that $\epsilon_{D}+\epsilon_{fluct}>0$ everywhere
translates into the requirement that the energy eigenvalue will
always be greater than the effective potential
$|\frac{\omega^2}{c^2}\epsilon_{fluct}({\bf x})|$. Therefore,
unlike the familiar picture of electronic localization, what we
are really seeking is an intermediate frequency window within the
positive energy continuum that lies at an energy higher than the
highest of the potential barriers. For those frequencies the
interference between the incident and the scattered EM waves will
be exactly destructive and this will allow no free propagating
photon modes to exist in the structure.

In the following section, we quantify these ideas in a more
rigorous way and present the procedure followed to find the
allowed modes as a function of the frequency, the {\it dispersion
relation}, in a simple 1-D periodic dielectric.
\begin{figure}
\centerline{\psfig{figure=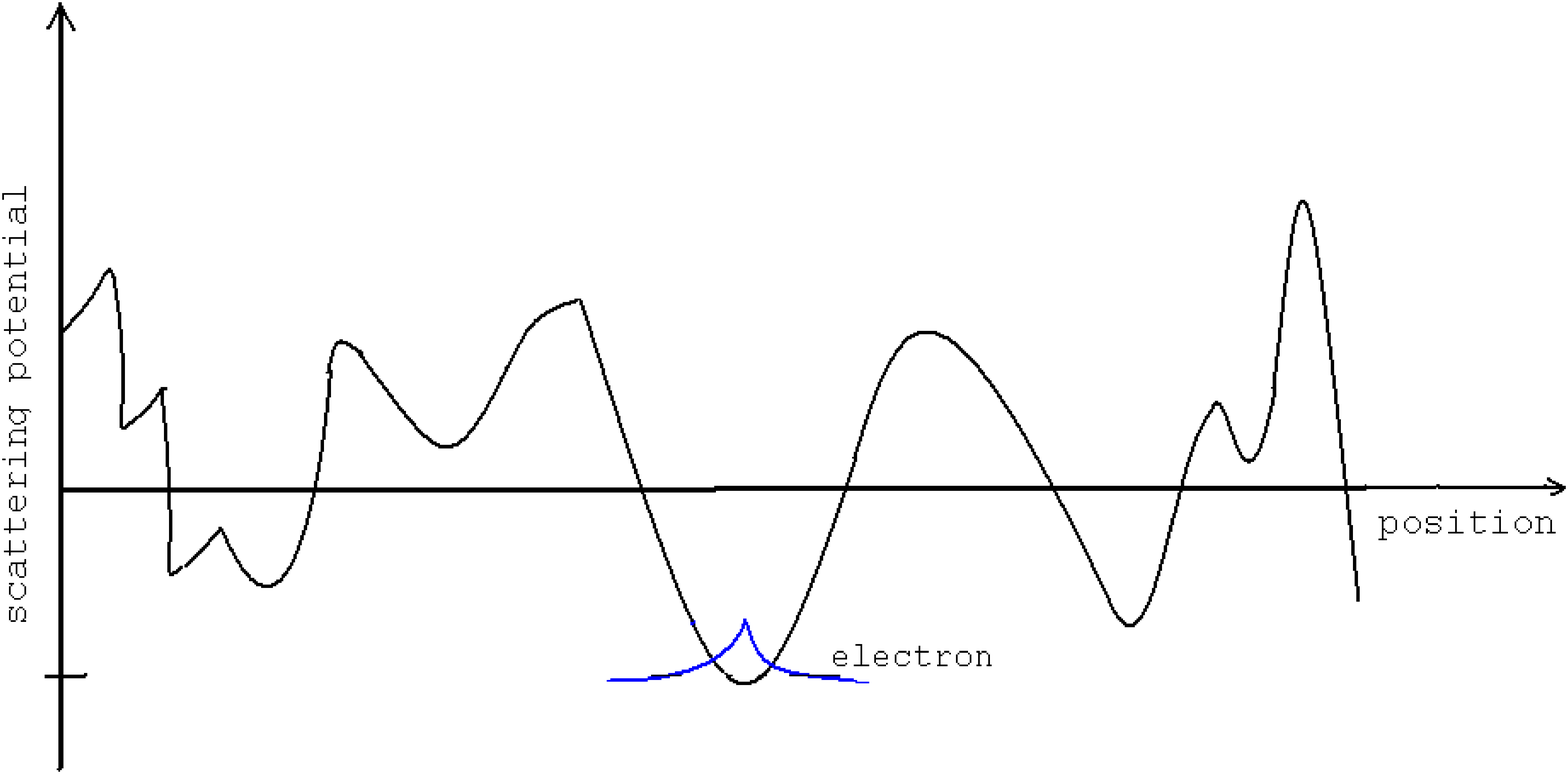,width=11.cm}}
\centerline{\psfig{figure=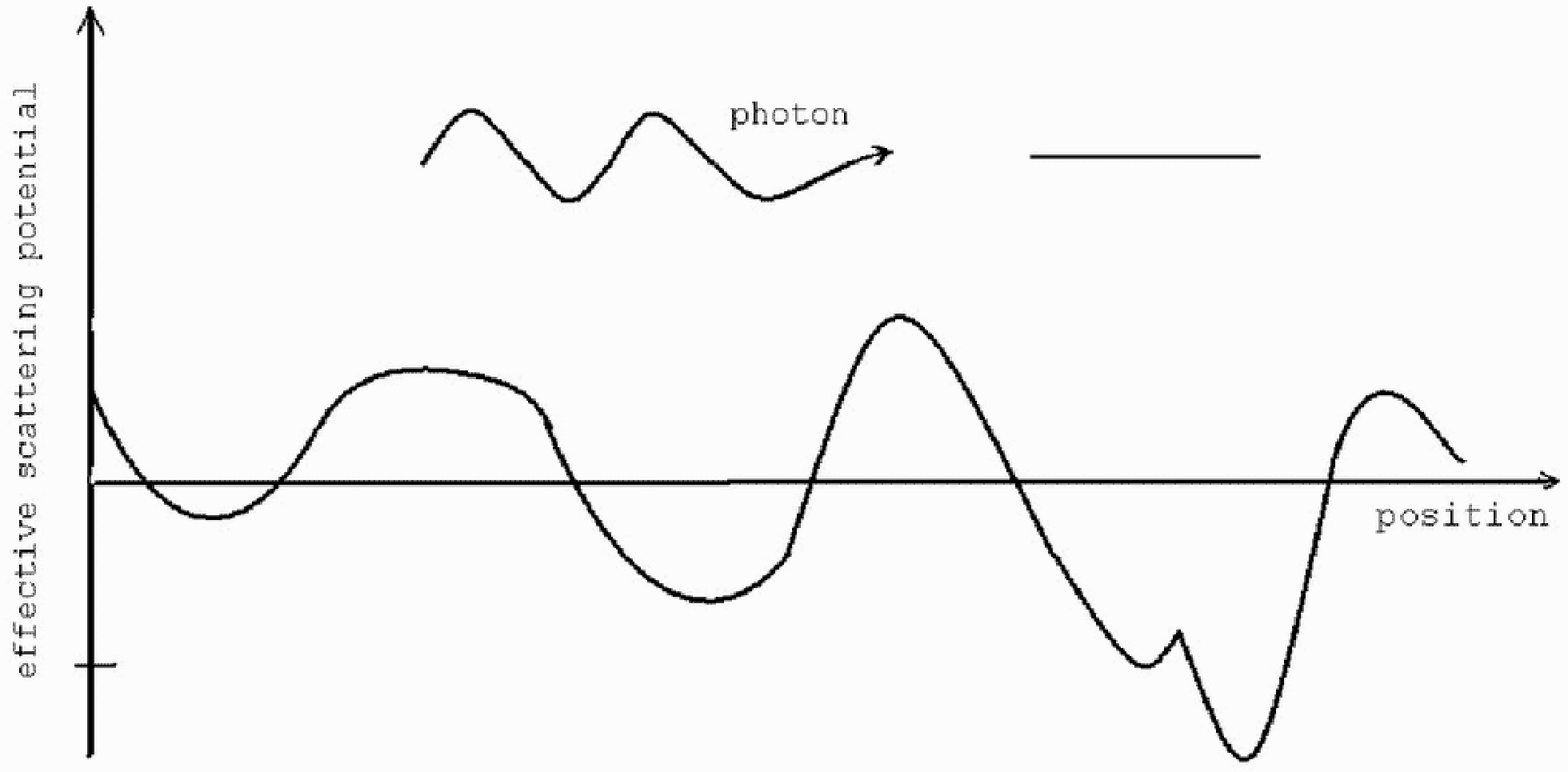,width=11.cm}}
\caption {The scattering potential for electrons in a solid (top
figure) and for photons in a random dielectric medium (bottom)
\cite{John91a}. The effective scattering potential for photons is
$(\omega^{2}/c^{2})\epsilon_{fluct}$, where $\epsilon_{fluct}$ is
the spatially varying part of the dielectric. The electron (top)
can have a negative energy, and can be trapped in deep potentials.
In contrast, the eigenvalue $(\omega^2/c^2)\epsilon_{D}$ of the
photon (bottom) must be greater than the highest of the potential
barriers if the dielectric constant is to be real and positive
everywhere. } \label{Fig3}
\end{figure}

\subsection{Calculating the dispersion relation}
\begin{figure}
\centerline{\psfig{figure=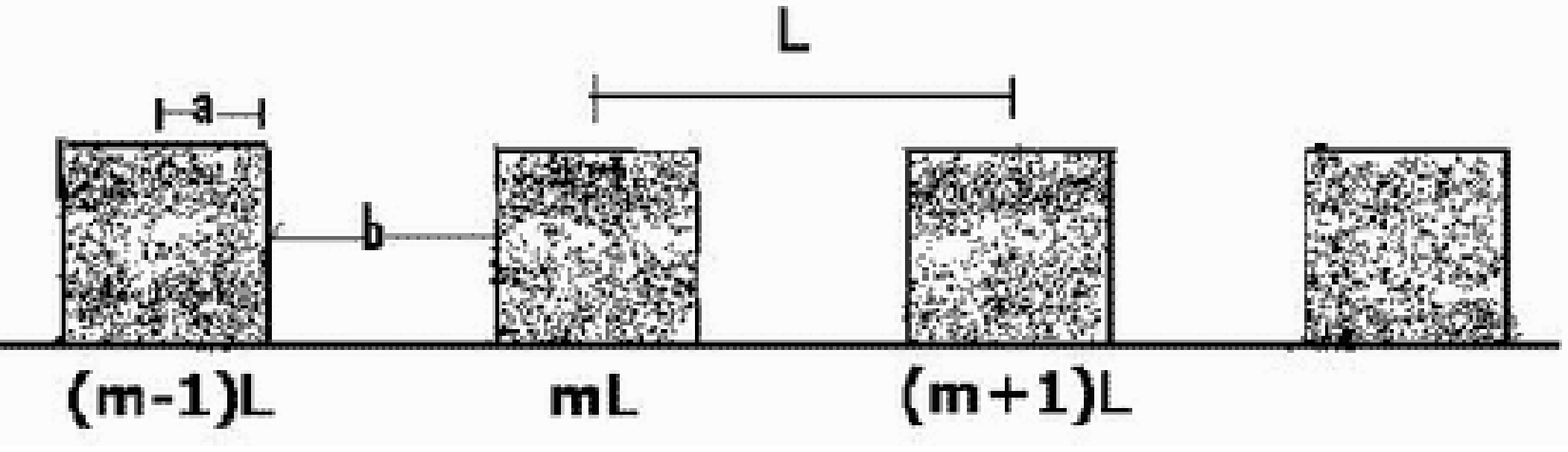,height=6.cm}} \caption
{The periodic potential. $L$ is the lattice constant}
\label{periodic}
\end{figure}

The dielectric constant of any photonic crystal can be expressed
as
\begin{eqnarray}
\epsilon({\bf x})=\epsilon_{D}+\epsilon_{fluct}({\bf x}) \, .
\end{eqnarray}
In this case we will assume that our structure exhibits
periodicity in one dimension and is homogenous in the other two.
More specifically
\begin{eqnarray}
\epsilon_{fluct}(x)=\sum_{n=-\infty}^{+\infty}u(x-nL) \, ,
\end{eqnarray}
with $L$ being the lattice constant and
\begin{eqnarray}
u(x)=\left\{
\begin{array}{lr}
n^2-1, &|x|<a \\ 0 ,&\mbox{otherwise} \end{array}\right. \, .
\end{eqnarray}
As discussed in the previous section, the propagation of a
monochromatic EM field in an inhomogeneous, non dissipative
dielectric medium is governed by the following equation
\begin{eqnarray}
-\nabla^{2}{\bf \cal{E}}+\nabla(\nabla \cdot {\bf \cal{E}}
)-\frac{\omega^2}{c^2}\epsilon_{fluct}({\bf x}){\bf \cal{E}}
=\epsilon_{D} \frac{\omega^2}{c^2}{\bf \cal{E}} \, ,
\end{eqnarray}
where $\cal{E}$ is the field's ampitude. Setting $\epsilon_{D}=1$,
the above equation reads ($\nabla\cdot {\cal E}=0$, free charge is
zero)
\begin{eqnarray}
-\nabla^{2}{\cal
E}(x)-\frac{\omega^{2}}{c^2}\epsilon_{fluct}(x){\cal
E}(x)=\frac{\omega^{2}}{c^2}{\cal E}(x) \, .
\end{eqnarray}
Setting $\Phi(x)=-\frac{\omega^2}{c^2}\epsilon_{fluct}(x)$ we get
\begin{eqnarray}
-\nabla^{2}{\cal E}(x)+\Phi(x){\cal
E}(x)=\frac{\omega^{2}}{c^2}{\cal E}(x)\label{propag_APB} \, ,
\end{eqnarray}
where the potential $\Phi(x)$ is basically a sequence of
``potential barriers" of width $2a$ (see Fig. \ref{periodic}). The
objective is to solve Eq.  (\ref{propag_APB}) for the potential
$\Phi(x)$

\begin{figure}
\begin{center}
\centerline{\hbox{\psfig{figure=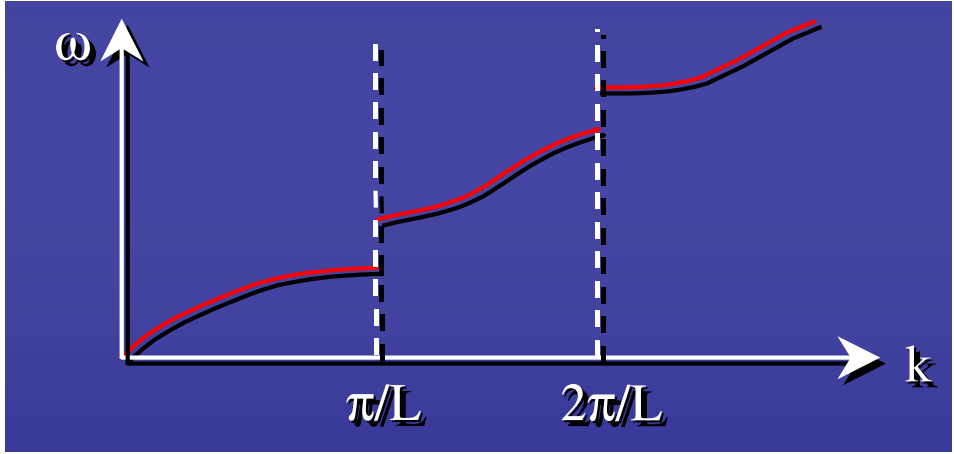,height=6.cm} }}
\end{center}
\caption {The dispersion relation for the 1-D isotropic model
shown in Fig. \ref{periodic}. As illustrated, gaps are formed at
$m\pi/L$ for integer values of $m$. } \label{periodic_disp}
\end{figure}

 Restricting ourselves to a unit cell of the
crystal, the solution can be expressed as follows
\begin{eqnarray}
{\cal E}(x)=\left\{ \begin{array}{lr} A e^{ikx}+B^{-ikx}, & x<-a \\
C e^{ik^{\prime}x}+De^{-ik^{\prime}x}, & |x|<a \, , \\
Ee^{ikx}+Fe^{-ikx}, &x>a
\end{array}
\right .
\end{eqnarray}
where $k=\omega/c$ and $k^{\prime}=n\omega/c$. According to the
Floquet theorem, ${\cal E}(x)$ should obey the following equation
\begin{eqnarray}
{\cal E}(x+L)=e^{ikL}{\cal E}(x),
\end{eqnarray}
and the derivative
\begin{eqnarray}
\frac{d{\cal E}(x+L)}{dx}=e^{ikL}\frac{d{\cal E}(x)}{dx} \ .
\end{eqnarray}
In addition the boundary conditions are that both ${\cal E}(x)$
and $d{\cal E}(x)/dx$ should be continuous at $x=\pm a$. Applying
all these conditions, we calculate the expansion coefficients and
also the following transcendental equation describing the relation
between $\omega$ and $k$:
\begin{eqnarray}
cos(kL)=cos(\frac{2na\omega}{c})cos(\frac{b\omega}{c})-
\frac{n^{2}+1}{2n}sin(\frac{2na\omega}{c})sin(\frac{b\omega}{c}) \
,
\end{eqnarray}
which for $b=2na$ can be inverted analytically providing the {\it
dispersion relation} for the 1D crystal,
\begin{eqnarray}
\omega_{k}=\frac{c}{4na}arccos[\frac{4ncos(kL)+(1-n^2)}{(1+n)^2}]
\, .
\end{eqnarray}
This dispersion relation leads to gaps at $k=\frac{m\pi}{2(n+1)a}$
for odd integer values of $m$ (see Fig. \ref{periodic_disp}). The
lowest gap is centered at the frequency $\omega_{gap}=\pi c/(4na)$
which for the case $b=2na$ equals $\pi/L$ (case shown Fig.
\ref{periodic}).

\begin{figure}
\centerline{\psfig{figure=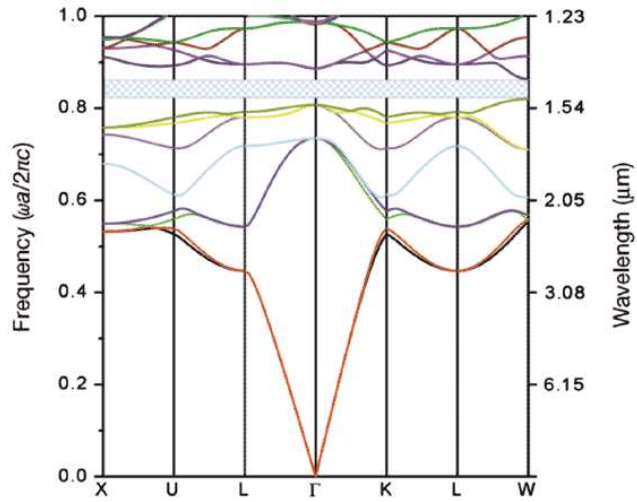,width=9.cm}} \caption {The
band structure calculation of a realistic 3-D photonic crystal(the
inverse opal structure of Fig. 1) according to theoretical
calculations. The shaded region corresponds to the gap predicted
(width  $5.1\%$) and partially observed by the Toronto group
\cite{Blanco00a}.
   }
 \label{opal_bands}
\end{figure}

In the more general case  of a realistic 3-D photonic crystal, the
usual approach is using the Bloch-Floquet theorem to expand both
the field amplitude and the dielectric constant in plane waves
whose wavevectors are reciprocal lattice vectors.
\begin{eqnarray}
{\bf \cal E}^{\bf k}_{\bf \omega}(\bf r)=\sum_{\bf{G},l}b_{{\bf
g},l}{\bf e}_{l}e^{i({\bf k+G})} \, ,
\end{eqnarray}
where ${\bf e}_{l}$ is the polarization vector and ${\bf G}$ are
reciprocal lattice vectors. Using this expansion in Eq.
(\ref{intro2}), the problem is reduced to the solution of a system
of linear equations on $b_{{\bf g},l}$ the solution of which
provides the allowed mode frequencies for a given crystal (Fig.
\ref{opal_bands}). We should note here that for more complicated
3-D structures the solution of the corresponding eigenvalue
equation proves to be cumbersome and highly efficient
computational techniques are usually required, see for example
Refs.\
\cite{Stefanou92a,Pendry96a,Stefanou98a,Ward99a,Stefanou00a,Modinos01a,Johnson01a}.
\subsection{Fabrication of photonic crystals}

In order to realize a 3-D structure with a full band gap for the
propagation of light for some specific frequencies, we need not
only to show that a specific geometry could in principle exhibit a
band gap  but also that the specific micro-structure is amenable
to microfabrication in the lab \cite{Koenderink02a}. To illustrate
the methods followed by various groups in this field we will start
with the simplest case, the traditional multilayer film. The
latter is the simplest dielectric structure where one can observe
inhibition of the linear propagation of a EM wave. It is
relatively easy to construct by assembling together dielectric
layers with alternating high and low refractive indexes.

\begin{figure}
\centerline{\psfig{figure=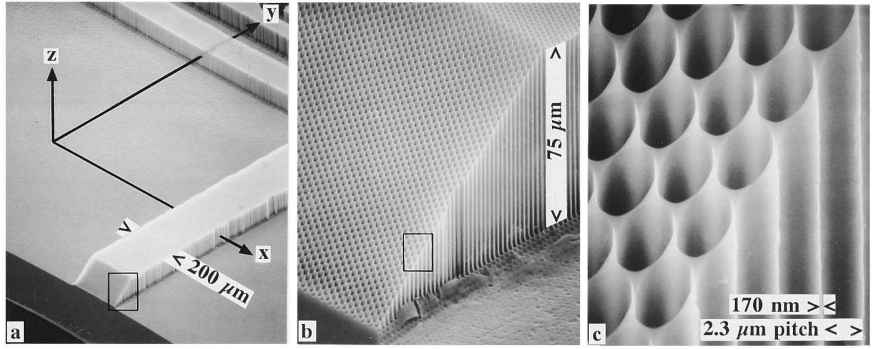,width=12.cm}}
\caption {Macroporous silicon forming a 2-dimensional triangular
lattice taken using scanning electron techniques
\cite{Gruning96a}. (a) A porous silicon bar of width 200$\mu m$
and height 75$\mu m$. (b) A tenfold magnification of the inset
shown in (a). (c) A tenfold magnification of the inset shown in
(b). The lattice constant is of the macropore array is 2.3$\mu m$,
the pore diameter 2.13$\mu m$, and the thinnest parts of the pore
walls 170nm.} \label{2D_triangular_exper}
\end{figure}

To go further than that, i.e. to create a band gap in the
propagation of an EM wave in two dimensions, we need something
more sophisticated than simply contrast in the index of
refraction. We need to find a specific geometry that will provide
a full band gap for a range of frequencies in two dimensions. For
this we need to keep in mind that the EM field consists of two
type of modes. The transverse-electric (TE) and transverse
magnetic (TM) ones. To achieve a complete band gap for all
polarizations, the corresponding bands should not only exist but
also overlap. TM band gaps are favored in a lattice of isolated
high index regions as in an array of dielectric columns in air. In
contrast, TE band gaps are favored in the {\it inverse} structure
as in an array of air columns (veins) drilled in a dielectric
substrate. Therefore to achieve a full band gap, we need somehow
to reconcile these seemingly contradictory conditions
\cite{Joannopoulosbook}. A structure satisfying this, is a
triangular lattice of low index columns (air) inside a high index
medium (silicon), see Fig. \ref{2D_triangular_exper}. More
specifically for the
 case where the radius of the columns is large enough, the
spots between the columns behave like localized regions of high
refractive index material and thus the above requirement should be
satisfied. It was predicted that for $r/a=0.48$ for the ratio
between the radius of the air columns to the lattice constant, a
complete photonic band gap should form exhibiting a gap-midgap
frequency ratio of about 19$\%$. The experimental verification was
provided by Gruning et el. \cite{Gruning96a}, who constructed the
crystal using an electrochemical technique to etch out columns  in
a silicon substrate. Their subsequent measurements verified a band
gap at $\lambda=5\mu m$. Also at the same year Krauss et al.
constructed 2D polarization-sensitive photonic band-gaps with
wavelengths in the range 800 - 900 nm \cite{Krauss96}.

\begin{figure}[t]
\centerline{\psfig{figure=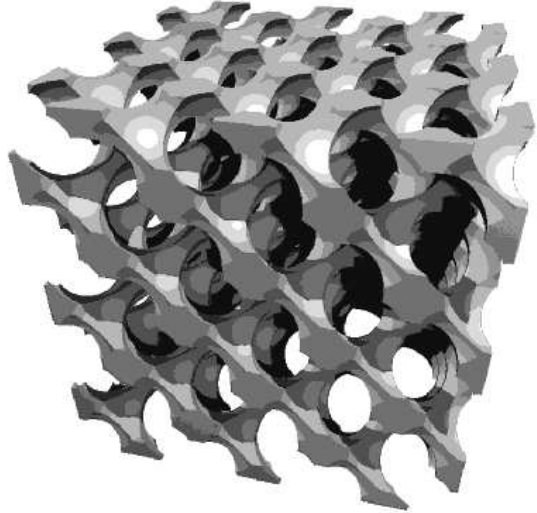,width=8.cm}} \caption { The
first prototype structure predicted to exhibit a large and robust
3D PBG structure designed by Yablonovitch \cite{Yablonovitch91a}.
It consists of an overlapping array of air spheres arranged in a
diamond lattice. It was created by drilling an array of
criss-crossing cylindrical holes into a bulk dielectric of
refractive index $n=3.6$ where band gaps of the order of 20$\%$
where demonstrated. (Picture courtesy of the Toronto group)}

\label{diammond}
\end{figure}

However, to achieve complete inhibition of light in all
directions, we need a 3D structure which exhibits periodicity in
all three directions. Out of the plethora of different geometries,
the one that was found, in theory first \cite{Ho90a} and
experimentally later \cite{Yablonovitch91a} to support a full 3-D
band gap is the {\it diamond} lattice \footnote{We note here that
proposed common semiconductors such as silicon and germanium also
follow diamond geometry.}. The structure initially suggested by
the Iowa group consists of either dielectric spheres in air or air
spheres embedded in a dielectric medium.
 Yablonovitch managed to implement the latter by mechanically
  drilling cylindrical holes
through a dielectric block (with $n\sim3.6$). The points where his
tunnels coincided formed a diamond-like structure (see Fig.
\ref{diammond}). In spite of the apparent
 simplicity and initial success of the method in
demonstrating the existence of a band gap in the microwave region,
 the application of the same method for optical waves proved to be
of great difficulty. In this case, the spheres and consequently
the ${\it drill}$ had to be in the micrometer region! We remind
here that in order for the TM and TE
 gaps to overlap, both constituent materials (air and dielectric)
 have to be topologically
interconnected and also a large contrast (at least 3) between the
two is essential. These constraints limited the use of
microengineering fabrication techniques (electron beam and X-ray
 lithography \cite{Lin98a} which proved to be very successful in longer wavelengths
\footnote{We note here the {\it ion drill} proposed by the
Yablonovitch group which led to the fabrication of structures with
band gaps in the near-infrared (1.1-1.5$\mu m$). The problem in
this case was that only a few unit cells could be reliably
produced.}.

\begin{figure}[t]
\centerline{\psfig{figure=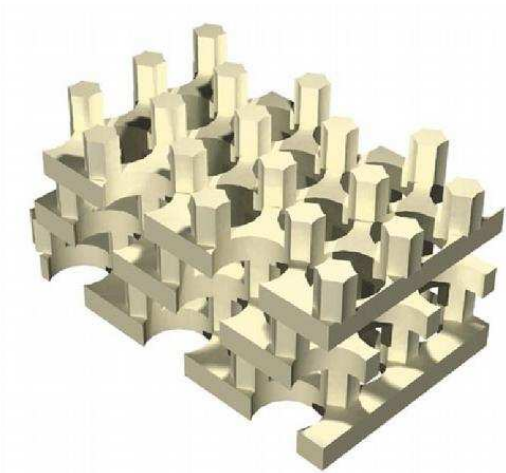,width=7.cm}} \caption {
Computer reproduction of a novel 3-D photonic crystal suggested by
the Joannopoulos group \cite{Johnson00a}. The structure could be
fabricated layer by layer following a three layer period thus
allowing the use of traditional lithographic techniques, along
with a high degree of control in placing defects (waveguides,
cavities, and other optical components) in the crystal. The layers
consist of alternating stacks of the two characteristic types of
2-D (or slab) photonic crystals: dielectric rods in air and air
holes in dielectric.
 }
\label{joan_layer}
\end{figure}

\begin{figure}
\centerline{\psfig{figure=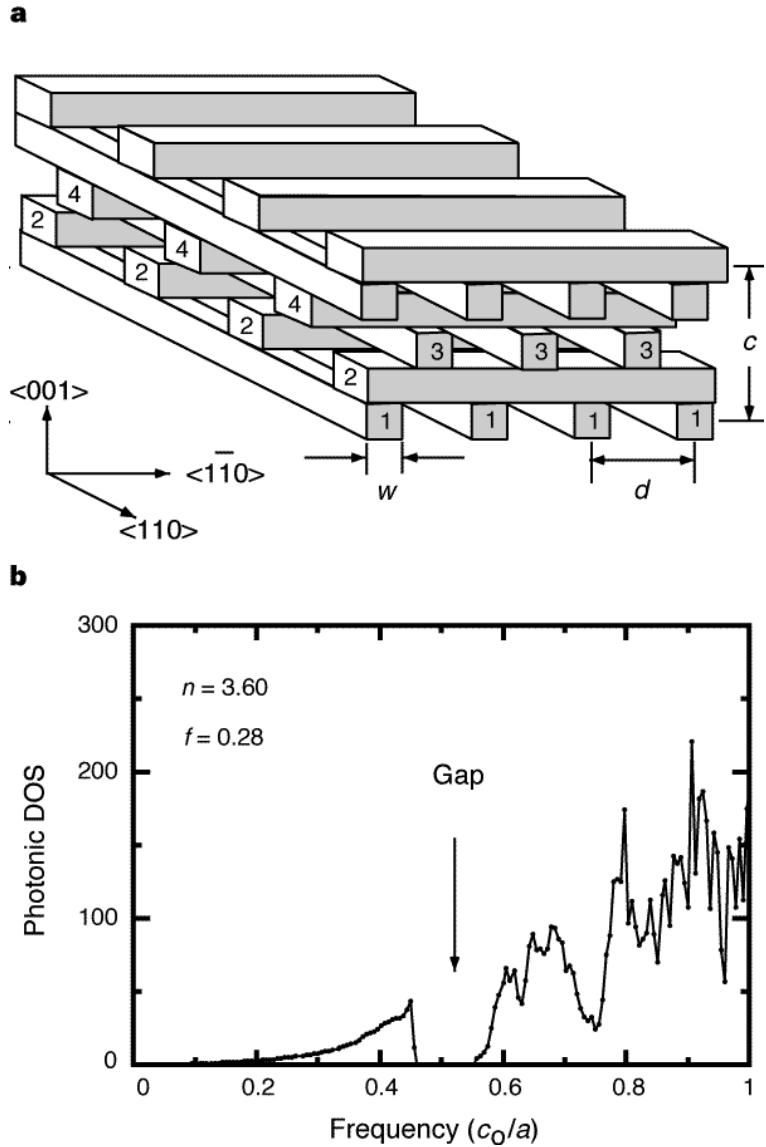,width=10.cm}}
\caption {The woodpile structure fabricated by repetitive
deposition and etching of multiple dielectric films. {\bf a},
Sketch of the layer-by-layer 3-D photonic band gap structure. It
is made from layers of one-dimensional rods with a stacking
sequence that repeats itself every four layers, a unit cell, with
a repeat distance of c. {\bf b}, Plot of the computer calculated
photonic density of states assuming that the refractive index of
the rods is n = 3.60, the filling fraction of the 3-D structure f
= w/d = 0.28, and c/d = 1.414. A complete photonic bandgap exists
from 0.46 c/ a to 0.56 c/ a, where c is the speed of light in
vacuum \cite{Lin98a}.
 }
\label{woodpiles}
\end{figure}

A different approach was the layer stacking technique proposed by
Ho {\it et al.} \cite{Ho94a} (see Fig. \ref{woodpiles}) and
implemented by Lin {\it et al.} \cite{Lin99a} where a gap at 1.5
microns was reported and the seven layer crystal exhibited 1$\%$
transmission (it was believed to drop to 0.1 $\%$ with ten
layers). Along similar lines using a combination of electron beam
lithography and reactive ion etching Noda {\it et al.} managed to
stack semiconductor rods with micrometer dimensions
\cite{Noda96a}. They reported 99.9 \% attenuation between 6 and 9
$\mu m$ using eight layers. We note here that both methods are
promising for any large scale technological application (see next
section)
 as they are characterized by low cost and high reliability.
\footnote{The MIT group led by Joannopoulos has recently proposed
to build
 a 3-D crystal by stacking 2-D ones (see Fig \ref{joan_layer}). They predict the
 opening of a full 3-D band gap.}

To overcome the difficulties of 3-D sub-micron engineering it has
also been proposed to utilize systems that tend to self assemble
themselves into various geometries. The most promising one proved
to be colloidal crystals and artificial opals.
 Colloidal particles have been synthesized by materials
such as latex and $SiO_{2}$ in the range of a few nanometres to a
few hundred micrometers. A suspension of colloidal microspheres,
with a typical concentration of $10^{10}$ particles/$cm^3$, can
residue under gravity into a cubic-closed-packed structure with
size of the order of 1cm. \footnote{Their crystalline structure
resembles that of natural opals.}

\begin{figure}
\centerline{\psfig{figure=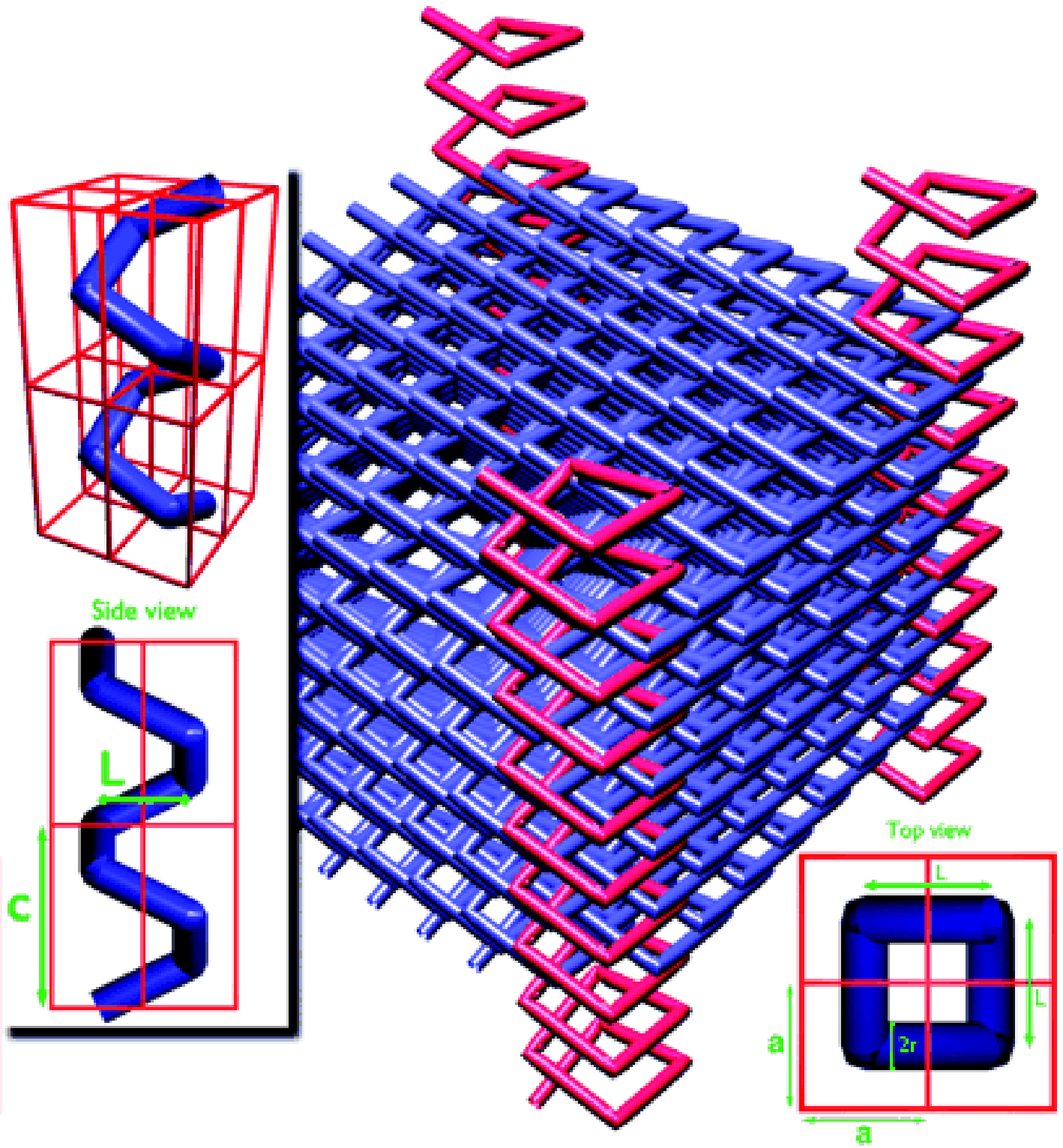,width=8.cm}}
\centerline{\psfig{figure=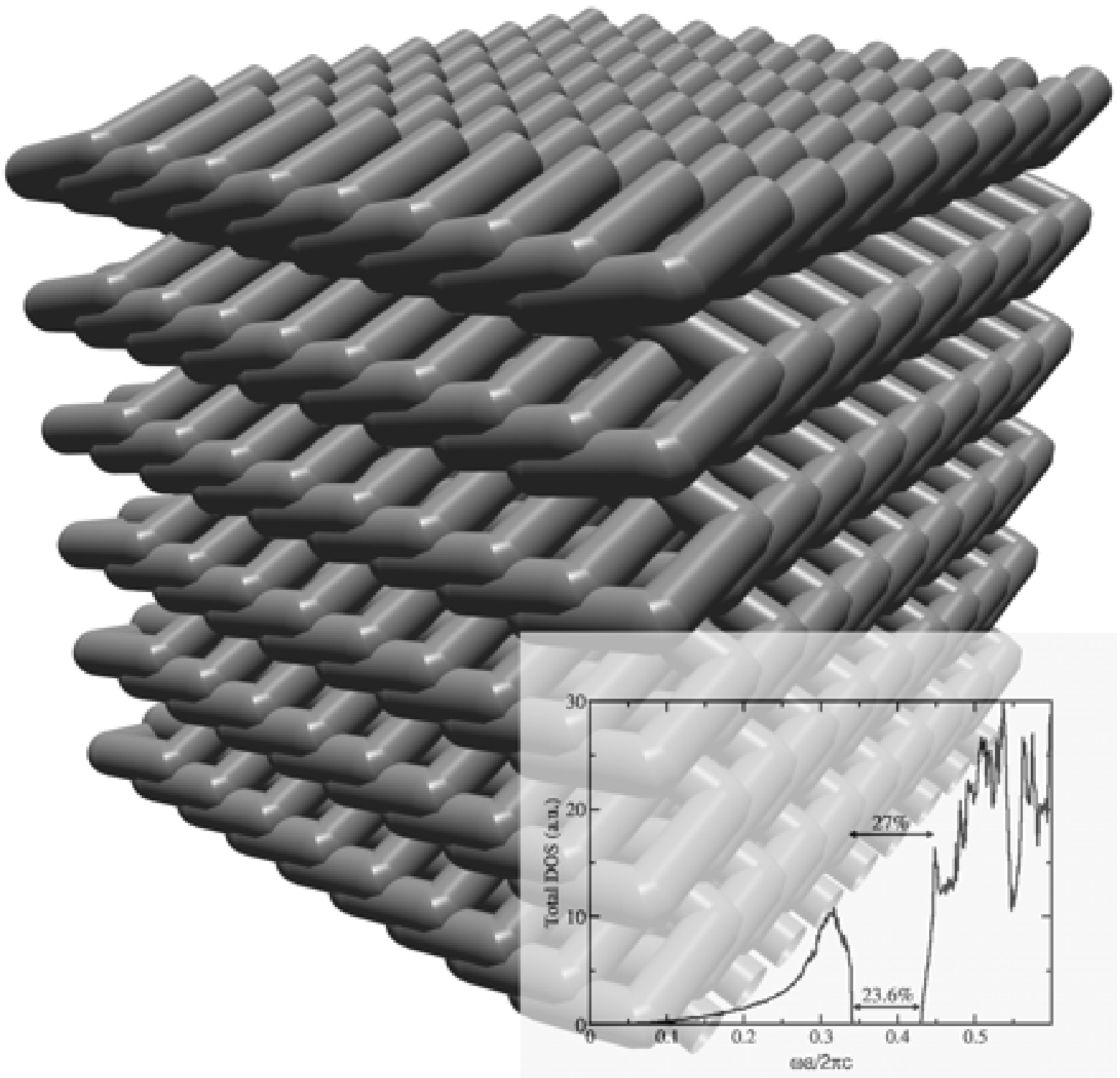,width=8.cm}}
\caption {(Top): The spiral crystal made of square spiral posts in
a tetragonal lattice proposed by Toader and John. It is believed
that structures like these might provide wider and more robust to
disorder band gaps \cite{Toader01a}. (Bottom): The Si crystal
obtained by inverting this template. A gap of 23.6\% occurs
between fourth and fifth bands of the photon dispersion relation.
The corresponding total DOS is shown in the inset (arbitrary
units). } \label{spirals}
\end{figure}

 In spite of the advantage of producing
inherently 3-D structures \cite{Winjhoven98a} compared to the 2-D
ones using lithographic techniques, it appears to be extremely
difficult to achieve the required refractive index contrast and
interconnectedness  for a full 3-D band-gap to open. On the other
hand, if they are used as templates to fabricate $inverse$ opals
(Fig \ref{John_inv_opal}), i.e., cubic-closed-packed lattices of
air {\it bubbles} on a dielectric matrix (silicon or GaAs)
\cite{Blanco00a,Schriemer01a,Galisteo03a}, near visible photonic
band gaps with gap to mid-gap frequency ratio of about 10$\%$ were
predicted (see Fig. \ref{opal_bands}). We note here that the void
regions that are left behind after the etching of the original
template (Swiss cheese structure) will allow the injection of
atoms or dye molecules thus making quantum optical type of
experiments possible.\footnote{In Ref. \cite{Angelakis02a} a
similar type of experiment for Bell Inequality test is proposed.}
In spite of the initial success of the inverted  opals in
providing rather large band gaps in infrared region, problems of
gap instability to disorder effects may prove to be difficult to
overcome. A solution to this was recently proposed  by S.John's
group in University of Toronto \cite{Toader01a}. They proposed an
alternative photonic crystal architecture consisting of square
spiral posts in a tetragonal lattice (Fig. \ref{spirals}). It
seems that this structure could exhibit a quite large and robust
3D PBG occurring between the fourth and fifth EM bands and is also
amenable to large-scale microfabrication using glancing angle
deposition (GLAD) techniques. Spiral post lattices with microscale
features have previously been synthesized using the GLAD method.
In this technique, complex 3D structures can be fabricated by
combining oblique vapor deposition and precisely controlled motion
of a two-dimensionally patterned substrate. In addition, their
square spiral posts can serve as templates for growing PBG
materials from an even larger range of materials. In this case, a
high refractive index material may be infiltrated to fill the void
regions between the posts, with the posts subsequently removed by
some selective etching process, leaving behind an ``inverted
structure." Their calculations have shown that the new inverted
structure could exhibit an even bigger band gap which for the case
that the initial structure is infiltrated by silicon gaps, could
be of the order of 27$\%$.

\subsection{Applications}

\begin{figure}
\centerline{\psfig{figure=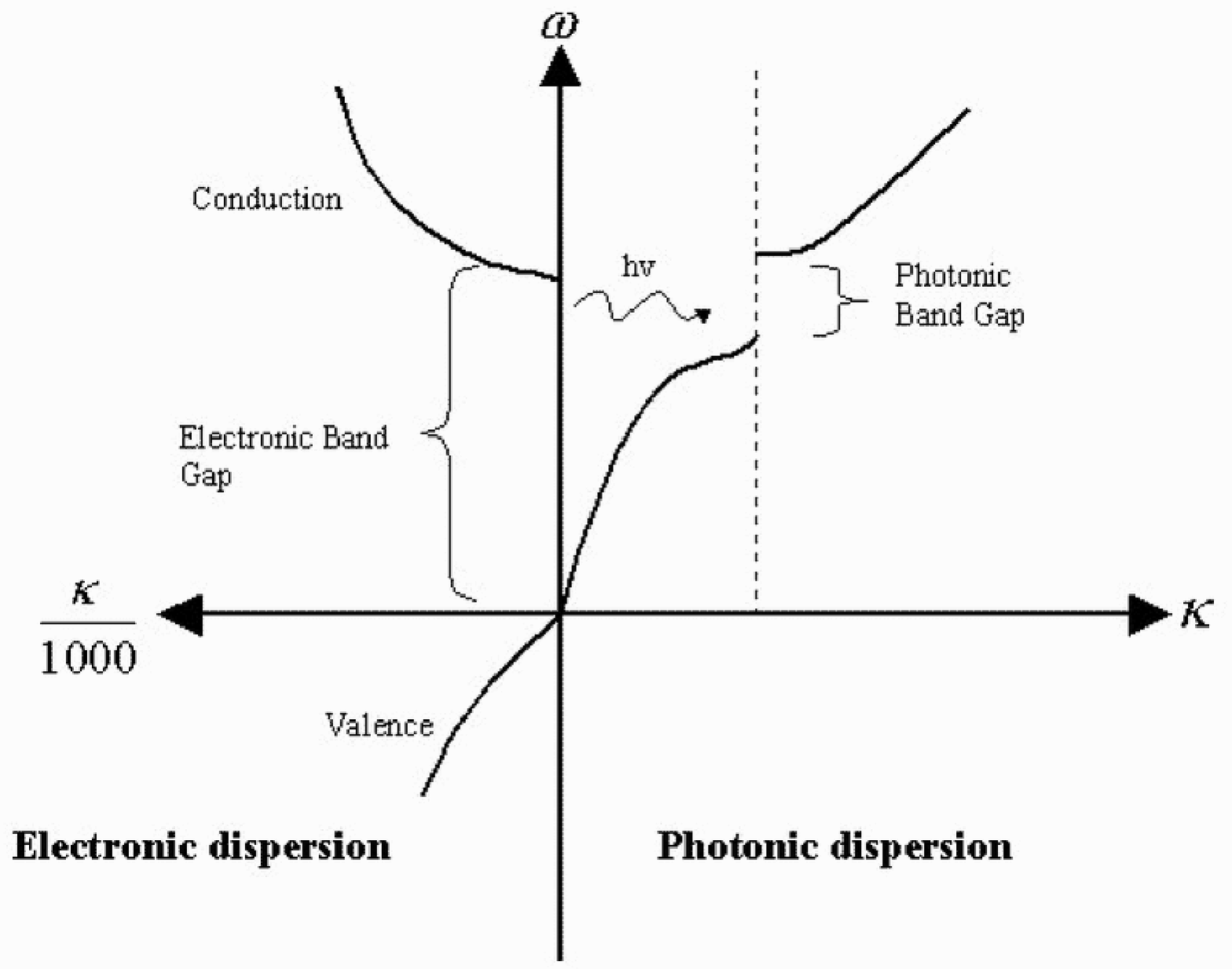,width=20.cm}} \caption {
Diagram showing the dispersion relation for both the photon and
electronic modes in a semiconductor integrated with a PBG
material. The EM dispersion relation for the PBG structure is
shown on the right whereas on the left is the electron wave
dispersion relation typical of a direct gap semiconductor. If an
electron for the conduction band were to recombine with a hole
from the valence band and the PBG overlaps the electronic band
edge, the resulting photon would not be emitted as there are no
modes available there! This will result in the inhibition of the
electron-hole recombination rate, seriously improving the
performance of devices like semiconductor lasers as Yablovonitch
has first suggested \cite{Yablonovitch01a}.
 } \label{both_gaps}
\end{figure}

\begin{figure}
\centerline{\psfig{figure=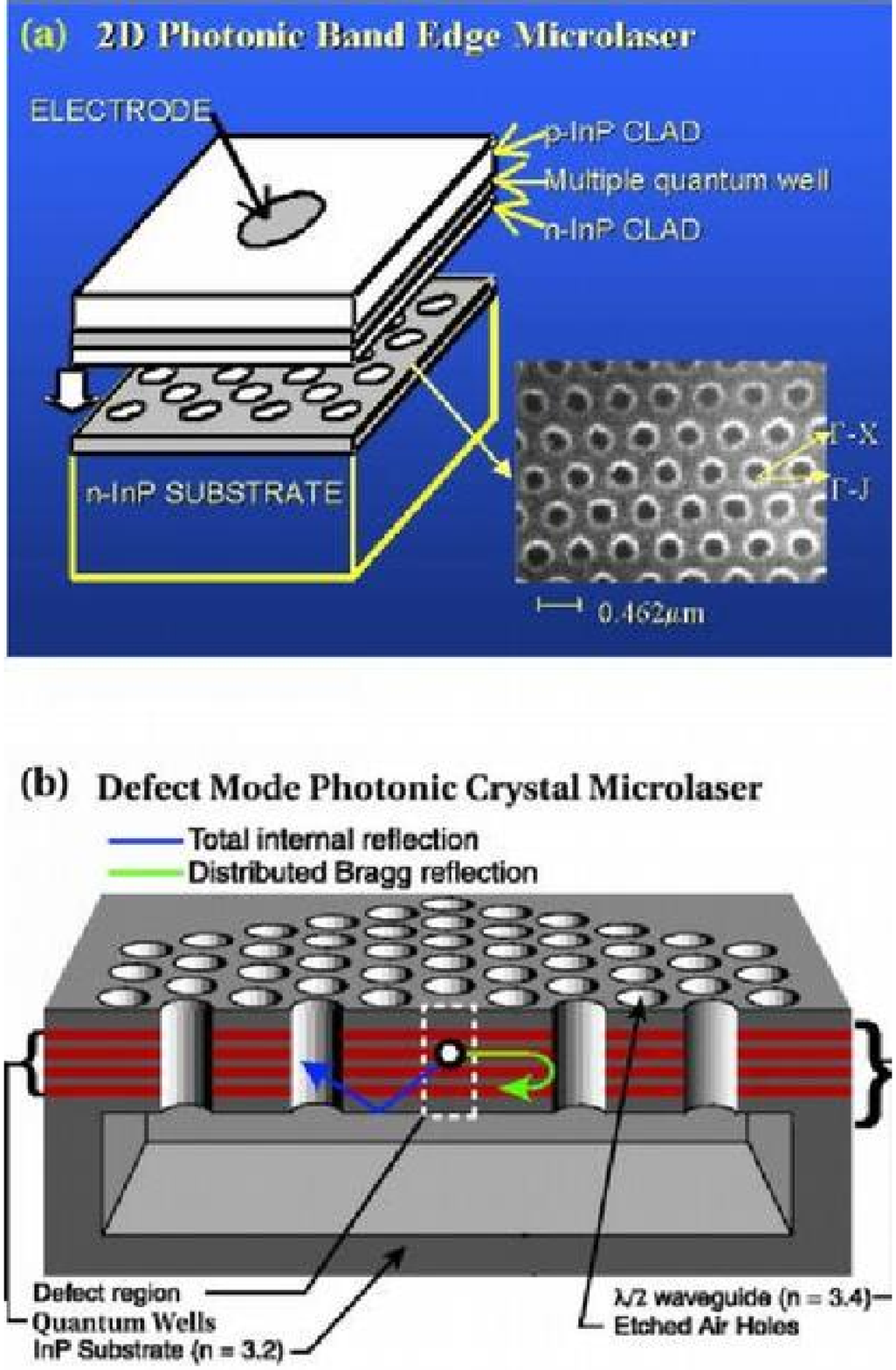,width=9.cm}} \caption {
Designs for 2-D photonic crystal micro-lasers. (a) The Band Edge
micro-laser. Stimulated emission (arising from electron-hole
recombination) from the multiple quantum well active region occurs
preferentially at the band edge. It has been proved that strong
feedback and memory effects can arise in that case \cite{John95a}.
For a real 3-D case this would lead to lasing without a
conventional cavity \cite{John95c,Florescu02a}. The Noda group has
realized a precursor to this where broad frequency emission occurs
preferentially at the band edge \cite{Imada99a} (courtesy of S.
Noda, Kyoto University). (b) Defect Mode micro-laser where a
defect mode with a localized state of light within the 2-D PBG
structure is engineered by allowing for a missing pore in the 2-D
photonic crystal. Stimulated emission from the multiple quantum
well active region falls mainly into the localized mode (courtesy
of Axel Scherer, California Institute of Technology).
 }
\label{2D-lasers}
\end{figure}

One of the most immediate applications of photonic crystals was to
be in optoelectronic devices where unwanted spontaneous emission
affected their performance. On the left of Fig. \ref{both_gaps} we
show the electron dispersion relation for a direct gap
semiconductor. Assuming that is embedded in a photonic crystal
(whose photonic dispersion relation is shown on the right of the
same graph) it is easy to see that if the PBG overlaps the
electronic band edge then the electron-hole recombination rate
could be inhibited. That is simply because the potentially emitted
photon will have no place to go! In a semiconductor laser this
would lead to near unity efficiency into the lasing mode. This
theoretical idea was recently implemented \cite{Imada99a} in the
2-D band edge micro-laser in which lasing occurs preferentially at
the 2-D photonic band edge even though the emission from the
active region has a broad frequency distribution (Fig.
\ref{2D-lasers}).

Along the same lines, light extraction from LEDs would be more
efficient. The main problem there is that only the light emitted
in a narrow angle \footnote{The efficiency
 being $1/(2n^2)$ where n is the index of refraction for the emitting material.
 For GaAs, n=3.5 and the efficiency is approximately 5 $\%$.} manages to escape
and the rest is usually trapped within the film and is either
absorbed or emerges from the edge of the device. There are several
ways to partially overcome this, for example using photon
recycling, rough surface, e.t.c., which could increase the
efficiency to 30$\%$ but none of them actually alters directly the
spontaneous emission properties of the device. If a 2-D photonic
crystal pattern in the form of a triangular lattice of air holes
is introduced into the semiconductor \cite{Fan97a}, extraction
efficiency of the order of $100\%$ is
 expected. This is simply due to the large
 band gap that will appear for in plane
propagation forcing the photons to come out of the slab in the
vertical directions. In addition, unlike the planar microcavity,
the efficiency is enhanced over a wide range of frequencies and as
no resonance or photon recycling is needed, the photon lifetime is
shorter. The latter of course leads to a reduction of the
absorption losses and an increase of the modulation speed of the
LED.

We proceed  by describing the potential applications of the
presence of {\it defect modes} in a photonic crystal. It was shown
that by  adding or removing a piece of dielectric material the
periodicity of the lattice is locally altered and this can lead to
the appearance of highly localized modes of light in an otherwise
mode free surrounding. In particular, removing a small amount of
high index material from one unit cell (air defect), leads to the
occurrence of a localized mode just above the top of the lower
band in analogy to the acceptor modes in semiconductors.
 On the other hand adding a small amount of high index material
  to a single unit cell (dielectric
defect) forces a single localized mode to split off from the upper
band edge as in semiconductor donor modes. The former case-an air
defect-is basically a high-Q microcavity with tunable frequency
\footnote{The frequency of the defect mode is an increasing
function of the volume of the defect area \cite{Joannopoulosbook}
and its energy is exponentially localized (usually within a few
lattice constants).}. All types of applications that involve
high-Q optical microcavities could thus implemented. The main
advantage over usual high-Q superconducting metallic cavities is
the operation on higher frequencies without almost any
losses\footnote{ It was experimentally shown \cite{Villeneuve96a}
that the quality factor increases {\it exponentially} with the
size of the crystal.}. Various devices such as frequency filters,
atom masers, zero-threshold lasers should be more efficient (see
Fig. \ref{2D-lasers}b). Also experiments involving single
atom-photon interaction in the strong coupling regime is important
should be easier realizable.

\begin{figure}
\centerline{\psfig{figure=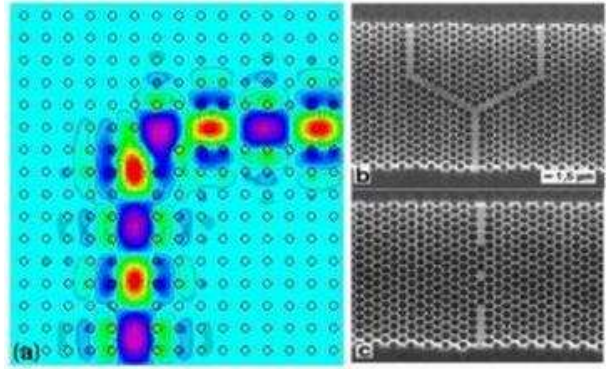,width=8.cm}} \caption {
Waveguiding in photonic crystals. (a) Right angle waveguide
channel in a 2D photonic crystal. The propagation of the field
modes is through a line a defect with no reflection or scattering
losses (courtesy of J.D.Joannopoulos group, Massachusetts
Institute of Technology). (b, c) Other line defect microstructures
fabricated in macroporous silicon with a lattice constant of 1.5
mm. The splitting of the modes in (b) could have application in
building optical devices such as interferometers. In (c), the air
holes within the line defect operate as reflectors or mirrors thus
creating a resonator cavity within the waveguide (courtesy of
Max-Planck Institute for Microstructure Physics, Halle, Germany).
} \label{waveguiding}
\end{figure}

In a similar way to the introduction of points defects, someone
can also create {\it line defects} in an otherwise perfect
structure. Such defects could be used as lossless waveguides
\cite{Mekis96a,Stefanou98b,Yariv99a,Fan01a,Yannopapas02a} where
light could be guided around sharp corners with no reflection or
scattering losses(see Fig. \ref{waveguiding}). Also highly
efficient fibers confining light within a hollow core (a large air
hole) in a silica-air photonic crystal fiber has been
demonstrated\cite{Cregan99}.

\begin{figure}
\centerline{\psfig{figure=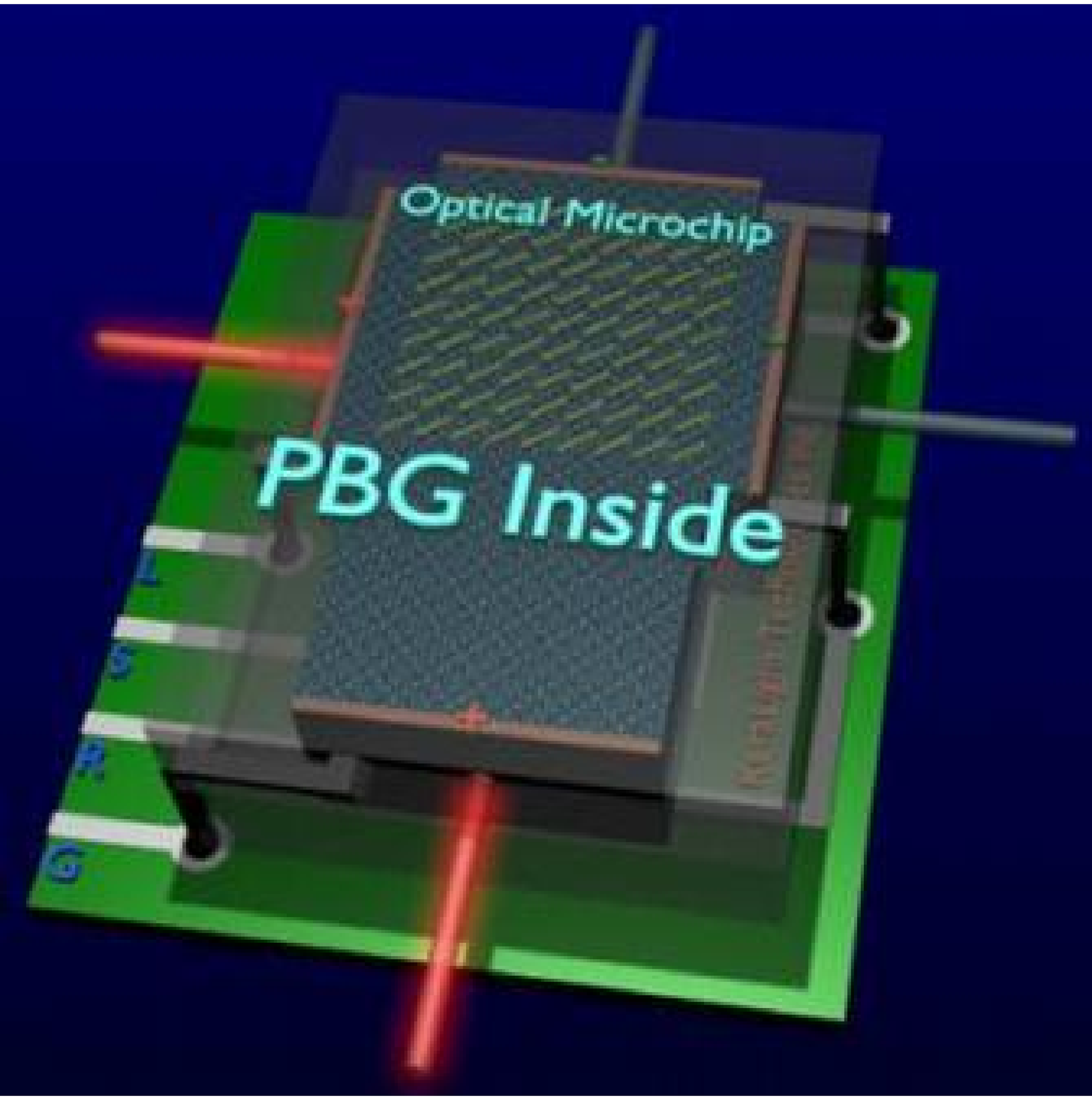,width=12.cm}} \caption {A
futuristic depiction of an electro-actively tunable PBG routing
device. The PBG structure has been infiltrated with an optically
anisotropic material such as a liquid crystal (indicated by the
yellow arrows) which responds strongly to external electric
fields. Applying a voltage, could alter the optical properties of
the whole structure opening or closing of the corresponding
photonic band gap could be achieved. This could lead to the
possibility of routing light from an optical fiber into one of
several output fibers (courtesy of S. John's group, University of
Toronto).
 }
\label{chip}
\end{figure}

Last but not least, we mention the idea of creating tunable PBGs
by the infiltration of inverted opal or spiral structures with
some low index of refraction liquid crystal. The potential of
controlling the direction of the nematic liquid crystal molecules
through an external electric field, could alter the optical
properties of the whole structure and simply provide external
control over the width of a full 3-D gap \cite{Busch99a}. The
resulting tunability of spontaneous emission, waveguiding effects,
and light localization may considerably enhance the technological
value of a composite liquid crystal PBG material over and above
that of either a bulk liquid crystal or a conventional PBG
material by itself. A tunable optical microchip which routes light
from a set of optical fibers and could be used as part of greater
optical circuit is shown in Fig. \ref{chip}.

We conclude this part by noting that the intense interest in the
field of photonic crystals in recent years has not solely been due
the above described ``optical circuit" type of applications, where
light was treated as a classical field. A lot of research has also
been towards the understanding of the light-matter interactions
within a photonic crystal at the quantum level. Several novel
phenomena, such as, for example, the inhibition of spontaneous
emission and the creation of ``atom-photon'' bound states
\cite{John90a,John91b,John94a,Kofman94a,Bay97a,Zhu97a,Bay97b,Quang97a,
Vats98a,Woldeyohannes99a,Nikolopoulos99a,Yang99a,Angelakis99a,Lambropoulos00a,Nikolopoulos00a,Zhu00a,
Yang00a,Yang00b,Zhu00d,Li00a,Li01a,Florescu01a,Wang02a,Zhang02a,Zhang03a,Woldeyohannes03a,Yang03a,Wang03a},
and also the existence of transparency near a photonic band edge
\cite{Paspalakis99a,Angelakis00a,Angelakis01a,Petrosyan00a,Du03a}
were found in these studies. Describing the effect of inhibition
of spontaneous emission and the creation of ``atom-photon'' bound
states will be the task of the section to follow.

\section{Spontaneous emission}

\subsection{The influence of the reservoir}

It was not until the 1980's \cite{Kleppner81a} before it was
established that spontaneous emission is not an intrinsic property
of matter over which we have no control on \cite{Weisskopf30a} but
a process greatly dependent on the nature of the surrounding
environment. We should note here that it was actually 1946 when an
effectively overlooked paper by Purcell \cite{Purcell46a} already
had suggested that the spontaneous emission rate of radiating
dipoles can be tailored by using a cavity to modify  the
dipole-field coupling and the density of the available photon
modes. If the modal density in the vicinity of the frequency of
interest is greater than that of free space, then spontaneous
emission will be enhanced, if it is less, it will be inhibited
This is the {\it Purcell effect}. Technological developments, in
the form of high quality and finesse cavities, extended this to
what we know today as cavity QED \cite{Haroche89a}. One of the
most characteristic concepts that  was introduced at the time was
the vacuum Rabi oscillations. To illustrate this, assume you have
an atom, excited to some Rydberg state, placed inside a microwave
cavity of high quality factor of the order of $10^{10}$ and mode
spacing larger than the mode width. Assuming that the frequency of
the transition from the initial Rydberg state to the nearest one
matches one of the cavity modes, then spontaneous emission is
greatly enhanced, whereas it is severely inhibited when the atomic
transition is detuned from all cavity modes by an amount larger
than the mode width. In the case of exact resonance, the
spontaneously emitted photon is reabsorbed by the atom then
reemitted again and so on, leading  to an oscillatory exchange of
energy between the atom and the cavity radiation field. The
operation of the micromaser is based on this effect where a
constant exchange of coherent microwave photons between the
mirrors of a microwave cavity and Rydberg atoms flying through
\cite{Hanch99a}.

From the implementation point of view, the technological expertise
required to built high-Q cavities has only recently moved from the
microwave to the optical or near optical regime allowing for the
possibility of similar research in the optical range
\cite{Yamamoto93a}. The great obstacle in localizing light in
cavities is related with the shape of cavity modes. Basically,
there are always tails of the usual Lorenzian distribution of the
cavity lineshape that extend to infinity allowing an excited atom
located in the cavity to eventually decay to the ground state.

 Extensive research in dielectric
structures of the form described in the previous section showed
that it was possible to fabricate situations where the modal
density of the electromagnetic field exhibited gaps for a range of
frequencies. As a qualitative approximation we would say that an
excited atom tuned with the gap should never decay as there are no
modes for its photon to exist.
 Initially lead by John and collaborators and then quickly
spreading to many groups throughout the word, an enormous amount
of work has been produced dealing with questions usually asked in
quantum optics, now extrapolated to PBGs, at least at their
theoretical properties under somewhat idealized situations using
generalized density of modes for the PBG material
\cite{John90a,John91b,John94a,Kofman94a,Bay97a,Zhu97a,Bay97b,Quang97a,
Vats98a,Woldeyohannes99a,Nikolopoulos99a,Yang99a,Angelakis99a,Lambropoulos00a,Nikolopoulos00a,Zhu00a,
Yang00a,Yang00b,Zhu00d,Florescu01a,Zhang02a,Zhang03a,Woldeyohannes03a,Yang03a}
or under more involved situations where the local density of modes
is used for the description of the PBG material
\cite{Li00a,Li01a,Wang02a,Wang03a}. The purpose of the sections to
follow is to review some of the tools used in the above studies
and more specifically in describing the dynamics of spontaneous
emission. The first part describes the case where the atoms
interact with the free space vacuum and the second with structured
continuum of a photonic crystal.

\subsection{Two-level atom in free space: Weisskopf-Wigner theory}

Suppose now we initially prepare our two-level atom in the upper
state of the doublet and allow it to interact with the modes of
the vacuum. The corresponding Hamiltonian, in the dipole and
rotating wave approximations, is given by
\cite{Scullybook,Meystrebook}
\begin{eqnarray}
H =\hbar\omega_{1}|1\rangle \langle1| + \hbar\omega_{2}|2\rangle
\langle2| + \hbar \sum_{k}\omega_{k}\beta^{\dagger}_{k}\beta_{k} +
\hbar\sum_{k}(|2\rangle \langle 1| \beta_{k} g_{21}(\omega_{k}) +
h.c.) \ , \label{ham}
\end{eqnarray}
where
\begin{eqnarray}
g_{21}(\omega_{k}) = -\frac{\omega_{12}\mu_{21}}{\hbar}
(\frac{\hbar}{2\epsilon_{0}\omega_{k}V})^{1/2}\hat{\epsilon}_{k}\hat{\mu}_{21}
\, ,
\end{eqnarray}
and
\begin{eqnarray}
 \hat{\mu}_{21}=\int\phi_{2}^{\*}({\bf x})e{\bf x}\phi_{1}({\bf x})d^3{\bf
 x} \, ,
\end{eqnarray}
is the dipole moment for the transition $|1\rangle$ (lower state
with wavefunction $\phi_{1}({\bf x})$) to $|2\rangle$ (upper state
with wavefunction $\phi_{2}({\bf x})$). Here, $\beta_{k}$ is the
annihilation operator for the field mode $k$, $V$ is the
quantization volume and $\epsilon_{0}$ is the electric
permittivity of free space. Also, $\hbar \omega_{j}$, $j = 1, 2$
is the energy of the $j$-th atomic level. Finally,
$\hbar\omega_{k}$ and $\hat{\epsilon}_{k}$ denotes the energy and
the polarization vector of the $k$-th reservoir mode.

Transforming the system in the interaction representation
\cite{Scullybook,Meystrebook} we obtain the interaction
Hamiltonian as
\begin{eqnarray}
H^{(I)} = \hbar\sum_{k}|2\rangle \langle 1| \beta_{k}
g_{21}(\omega_{k}) e^{-i(\omega_{k}-\omega_{12})t} + h.c. \ ,
\label{int_ham}
\end{eqnarray}
where $\hbar\omega_{12} = \hbar(\omega_{2} - \omega_{1})$.
Assuming that initially the field modes are in the vacuum state
$|0\rangle$,  the wavefunction of the system can be written in
terms of the state vectors as
\begin{eqnarray}
|\psi(t)\rangle = a_{2}(t)|2,0\rangle +
\sum_{k}b_{k}(t)|1,k\rangle \, , \label{wavefunction_intro}
\end{eqnarray}
where the coefficients  $b_{k}$ represent probability amplitudes
of emitting one photon of energy $\omega_k$ belonging to the
vacuum modes  $|1,k\rangle$ (with the atom in state $|1\rangle$).
Also $ a_{2}(t)$ represents the probability of the atom being in
the excited state and the field has no photons. The time evolution
of the system is described by the Shr\"{o}dinger equation
\begin{equation}
i \hbar \frac{\partial}{\partial t}|\psi(t)\rangle = H^{(I)}
|\psi(t)\rangle \, , \label{sch}
\end{equation}
which using Eq.  (\ref{int_ham}) leads to the following set of
first order, coupled linear differential equations
\begin{eqnarray}
i \dot{a}_{2}(t) &=& \sum_{k} g_{21}(\omega_{k})e^{-i\delta_{k}t}b_{k}(t) \, ,   \label{a20}  \\
i \dot{b}_{k}(t) &=&  g^{*}_{21}
(\omega_{k})e^{i\delta_{k}t}a_{2}(t) \, ,  \label{bk0}
\end{eqnarray}
where $ \delta_{k}=\omega_{k}-\omega_{12}$ is the detuning of the
emitted photon frequency from the atomic  transition. We formally
integrate Eq.  (\ref{bk0})  and obtain
\begin{eqnarray}
b_{k} = -ig^{*}_{21} (\omega_{k})\int^t_0 a_{2}(t^{\prime})
e^{i\delta_{k}t^{\prime}} \ dt^{\prime} +b_{k}(0) \, . \label{bk1}
\end{eqnarray}
Taking into account that $b_k(0) =0$,  we  substitute Eq.
(\ref{bk1})  into Eq. (\ref{a20}) and obtain
\begin{eqnarray}
\dot{a}_{2}(t) = -\int^t_0 dt{^\prime} a_{2}(t^{\prime})\ \sum_{k}
|g_{21}(\omega_{k})|^2 e^{-i\delta_{k}(t-t{^\prime})} \, ,
\label{a21}
\end{eqnarray}
which is still an exact equation. In order to proceed we have to
calculate the kernel
\begin{equation}
K(t-t^{\prime}) = \sum_{k} |g_{21}(\omega_{k})|^2
e^{-i\delta_{k}(t-t{^\prime})} \, . \label{kern1}
\end{equation}
The sum over modes  is generally, converted to an integral, by
including the density of modes (states) $ \rho(\omega_{k}) $,
\begin{eqnarray}
\sum_{k}|g_{21}(\omega_{k})|^2e^{-i\delta_{k}(t-t^{\prime})}=\frac{V}{(2\pi)^3}\sum_{\sigma}
\int
d\Omega\int_{0}^{\infty}|g_{21}(\omega_{k})|^2\rho(\omega_{k})e^{-i\delta_{k}(t-t^{\prime})}
, \label{sum_integral}
\end{eqnarray}
where $d\Omega$ is the solid angle and $\sigma$ the light
polarization.
 In free space, the product $|g_{21}(\omega_{k})|^2\rho(\omega)\sim\omega$ is a smooth varying
function of $\omega$. Taking this into account we can approximate
\begin{eqnarray}
\int d\Omega\int_{0}^{\infty}|g_{21}(\omega_{k})|^2\rho(\omega)
e^{-i(\omega_{k}-\omega_{21})(t-t^{\prime})}\approx\\
|g_{21}(\omega_{21})|^2\rho(\omega_{21})\int
d\Omega\int_{0}^{\infty}e^{-i(\omega_{k}-\omega_{21})(t-t^{\prime})}
\, ,
\end{eqnarray}
which eventually gives \cite{Scullybook,Meystrebook}
\begin{eqnarray}
K(t-t^{\prime})= \frac{\gamma_{21}}{2}\delta(t-t{^\prime})
,\label{kernel free}
\end{eqnarray}
where $\gamma_{12} =  {\omega_{12}}^3 {\mu_{21}}^2/(3 \pi
\epsilon_{0} \hbar c^{3})$ is the famous free space decay rate.
This approach is the much used the Weisskopf-Wigner approximation.

The fact that the so called response function $K(t-t^{\prime})$ is
a delta-function means that the free space acts as an immediate
response reservoir. In other words spontaneous emission is dealt
as a Markovian process and the evolution of the system depends
only on the present and not on any previous state of the
reservoir. Substituting Eq. (\ref{kernel free}) into (\ref{a21})
our initial set of equations become
\begin{eqnarray}
\dot{a}_{2}(t) &=& -\frac{\gamma_{21}}{2} a_{2}(t) \, , \\ i
\dot{b}_{k}(t) &=&  g^{*}_{21}
(\omega_{k})e^{i\delta_{k}t}a_{2}(t).
\end{eqnarray}
The above equations can easily be solved with respect to time
which gives
\begin{eqnarray}
|a_{2}(t)|^2=e^{-\gamma_{21}t} \, .
\end{eqnarray}


\subsection{Two-level atom in a photonic crystal}

\begin{figure}
\centerline{\hbox{\psfig{figure=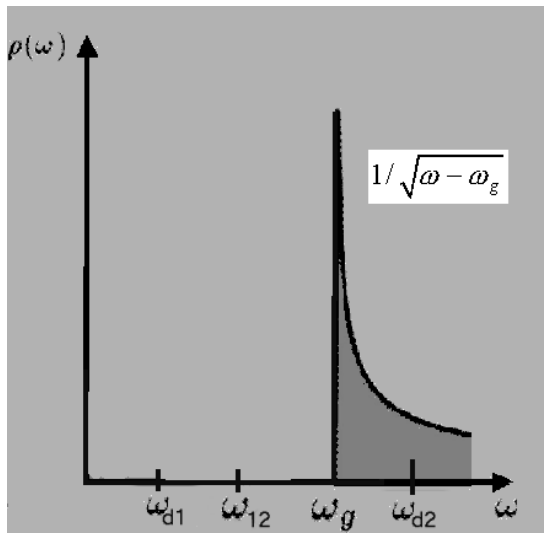,height=7.cm} }}
\caption {The density of states under consideration. $\omega_{12}$
is the atomic frequency and $\omega_{d1},\omega_{d2}$ the
corresponding dressed states frequencies. These states emerge due
to the ``self" dressing of the atom by its own localized radiation
field.} \label{dressed_in_gap}
\end{figure}

Assume now that our two-level atom is coupled to the radiation
field of a modified reservoir. We will specifically assume that
the atom is embedded in a three-dimensional photonic crystal where
the photon dispersion is found to be an isotropic one and
satisfies the following transcendental equation\footnote{We remind
 here that $n$ is the refractive index of the scatterer, $a$ is
its radius, and $2a+b=L$ is the lattice constant (see Section 2).}
\begin{eqnarray}
\omega_{k}=\frac{c}{4na}\arccos\left[\frac{4ncos(kL)+(1-n)^2}{(1+n)^2}\right]
.\label{disp}
\end{eqnarray}
The Hamiltonian and the wavefunction of the system are still given
from Eqs. (\ref{int_ham}),(\ref{wavefunction_intro}) and Eqs.
(\ref{a20}),(\ref{bk0}) still describe the evolution with the
difference now that atomic transition is occurring in the vicinity
of a band gap in the density of allowed photon modes inside the
photonic crystal.  As we showed in the previous section the
excited state amplitude $a_{2}$ is given by
\begin{eqnarray}
 \dot{a}_{2}(t) = -\int^t_0 a_{2}(t^{\prime})K(t-t{^\prime})\   \ dt{^\prime} \, . \label{k1}
\end{eqnarray}
In order to calculate the kernel in this case we cannot use the
Weisskopf-Wigner approximation as the density of states changes
very rapidly in the vicinity of the atomic transition when located
near the band edge.
 In this case, we must perform an exact integration in Eq.
  (\ref{kern1}). To do this, we observe that
in the vicinity of the gap the dispersion relation Eq.
(\ref{disp}) can be approximated as
\begin{eqnarray}
\omega_{k}=\omega_{g}+A(k-k_{0})^2 \, ,
\end{eqnarray}
where $A\simeq \omega_{g}/k_{0}^{2}$ (this corresponds to a
density of states of the form $\rho(\omega) \sim \Theta(\omega-
\omega_{g})/\sqrt{\omega - \omega_{g}}$ shown in Fig.
\ref{dressed_in_gap}, with $\Theta$ being the Heaviside step
function \cite{John90a}). The latter  combined with  Eq.
(\ref{sum_integral}) gives
 for the response function of Eq.  (\ref{kern1})
\begin{eqnarray}
K(t-t^{\prime})=\frac{\beta^{3/2}e^{-i[\pi/4+
\delta_{g}(t-t^{\prime})])}}{\sqrt{\pi(t-t^{\prime})}}, \quad t >
t^{\prime} \, , \label{kern2}
\end{eqnarray}
with $\beta^{3/2}=\omega_{12}^{7/2}|{\bf \mu}_{12}|^2/(6 \pi
\epsilon_{0} \hbar c^{3})$ and
$\delta_{g}=\omega_{g}-\omega_{12}$. Eq.\ (\ref{kern2})
demonstrates the non-Markovian character of the reservoir. In
contrast to the free space case we can see from Eq.\ (\ref{kern2})
that there is a contribution in the current dynamics at time $t$
from previous states of the system at time $t^{\prime}$ following
an inverse square root dependence. As we will discuss later on,
this is due to the partial localization of the emitted photon in
the vicinity of the atom where it can be re-absorbed and thus
affect the atom's evolution again after its initial emission.

We continue by trying to derive the explicit time dependence of
the atom's evolution. To do that we need to solve the
integro-differential equation (\ref{k1}). For this, we first
Laplace transform it  and obtain
\begin{eqnarray}
s{\tilde a}_2(s)-a_{2}(0)={\tilde K}(s) {\tilde a}_2(s) \, ,
\label{lapl}
\end{eqnarray}
where ${\tilde a}_2(s), {\tilde K}(s)$ are the Laplace transforms
of $a_{2}(t)$ and $K(t)$ accordingly. Using Eq.  (\ref{kern2}) we
obtain for the excited state amplitude ${\tilde a}_2(s)$
\begin{eqnarray}
{\tilde
a}_2(s)=\frac{(s-i\delta_{g})^{1/2}}{s(s-i\delta_{g})^{1/2}-(i\beta)^{3/2}}
\label{a_s}.
\end{eqnarray}

To obtain the dependence in the time domain we invert this using
the Bromwitch formula:
\begin{eqnarray}
a_{2}(t)=\frac{1}{2\pi
i}\int_{\epsilon-i\infty}^{\epsilon+i\infty}e^{st}{\tilde
a}_{2}(s)ds \, ,
\end{eqnarray}
where the real number $\epsilon$ is chosen such that $s=\epsilon$
lies to the right of all the singularities (poles and branch
points) of the function ${\tilde a}_{2}(s)$. The inverse Laplace
transform of Eq.  (\ref{a_s}) yields \cite{John94a}
\begin{eqnarray}
a_{2}(t)=2b_{1}x_{1}e^{\beta x_{1}^{2}+i\delta_{g}t}+
b_{2}(x_{2}+y_{2})e^{\beta x_{2}^{2}+i\delta_{g}t}- \nonumber \\
\sum_{j=1}^3 b_{j}y_{j}[1-\Phi(\sqrt{\beta x_{j}^{2}t})]e^{\beta
x_{j}^{2}+i\delta_{g}t}, \label{sol1}
\end{eqnarray}
where
\begin{eqnarray}
x_{1}&&=(A_{+}+A_{-})e^{i\pi/4},\label{x1}\\
x_{2}&&=(A_{+}e^{-i(\pi/6)}-A_{-}e^{i(\pi/6)})e^{-i(\pi/4)},\label{x2}\\
x_{3}&&=(A_{+}e^{i(\pi/6)}-A_{-}e^{-i(\pi/6)})e^{i(3\pi/4)},\label{x3}\\
A_{\pm}&&=\left[\frac{1}{2}\pm\frac{1}{2}\left[1+\frac{4\delta_{g}}
{27\beta^{3}}\right]^{1/2}\right]^{1/3},\label{A+-}\\
b_{j}&&=\frac{x_{j}}{(x_{j}-x_{i})(x_{j}-x_{k})} \qquad (j\ne i
\ne k; j,i,k=1,2,3),\\ y_{j}&&=\sqrt{x^2_{j}} \qquad (j=1,2,3),
\end{eqnarray}
and $\Phi(x)$ is the error function \cite{Gradshteynbook}. As we
see from Eq.  (\ref{x1}),  if $\delta_{g}=0$ then $\beta
x^2_{1}=i\beta$. This means that the value of $\beta$ given above
can be considered as a resonant frequency splitting, an analog of
the vacuum Rabi splitting in cavity quantum electrodynamics
\cite{Berman94a}. For large times, i.e. for large values of $\beta
t$, the terms of higher order than $(\beta t)^{3/2}$
\cite{Gradshteynbook} can be ignored, and Eq. (\ref{sol1}) reduces
to
\begin{eqnarray}
a_{2}(t)\cong 2b_{1}x_{1}e^{\beta x_{1}^{2}+i\delta_{g}t}+
b_{2}(x_{2}+y_{2})e^{\beta x_{2}^{2}+i\delta_{g}t}+
\frac{1}{2\sqrt{\pi}}\left[ \sum_{j=1}^{3}\frac{b_{j}}{x^{2}_{j}}
\right] \frac{e^{i\delta t}}{(\beta t)^{3/2}}\label{sol2}.
\end{eqnarray}

As we see from Eq. (\ref{sol1}) the atomic level splits into
dressed states caused by the atom and its own radiation field
located at the frequencies $\omega_{d1}=\omega_{g}-\beta
Im(x^2_{1})$ and $\omega_{d2}=\omega_{g}-\beta Im(x^2_{2})$. It is
easy to see using  (\ref{x1}) and (\ref{A+-}) that
$x_{1}^2=i|x_{1}|^2$.
 This translates to the fact the corresponding
dressed state at the frequency $\omega_{g}-\beta|x_{1}|^2$ is the
photon-atom bound dressed state with no-decay. A photon emitted by
the atom in this state will tunnel for a length scale of a few
lattice constants before being reflected back and re-absorbed by
the atom. The other dressed state on the contrary is pushed
outside the gap, where the density of photon states is not zero
and eventually will be responsible for the part of the initial
excitation that will eventually decay (see Fig.
\ref{dressed_in_gap}).
\begin{figure}
\centerline{\hbox{\psfig{figure=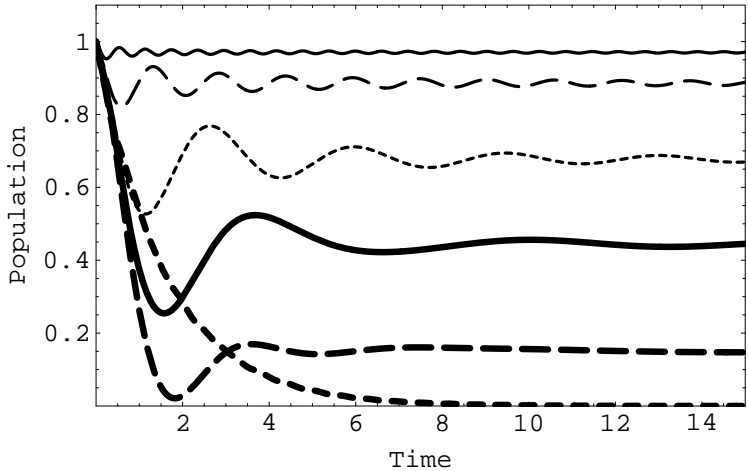,height=7.cm}
}} \caption {Atomic population on the excited state,
$P(t)=|a_{2}(t)|^2$ as function of time for various values of the
detuning from the band edge $\delta=-10\beta$ (thin solid curve),
$\delta=-3.5\beta$ (thin dashed curve),
 $\delta=-1\beta$ (thin dotted curve), $\delta=0$ (thick solid curve),
 $\delta=1\beta$ (thick long-dashed curve), $\delta=10\beta$ (thick short-dashed curve). The time is in units
 of $1/\beta$.}\label{pop_isotropic}
\end{figure}
In Fig. \ref{pop_isotropic} we show the atomic population as a
function of time for various detunings from the band edge
frequency. As was expected for atomic transitions well inside the
gap ($\delta_{g}=-10\beta$), the atom remains in the excited state
for ever. In this case the second term in Eq.\ (\ref{sol1})
vanishes which means that no true atomic level splitting is
present in this case and the oscillations in the dynamics are
caused by the interference with a ``quasidressed'' state
originating from the branch point term.  As $\delta_{g}$ moves
from negative to positive, the other component, located at
$\omega_{g}-\beta|x_{1}|^2$ becomes important and eventually for
$\delta_{g}=10\beta$ is the major one forcing the atom to complete
decay in the usual exponential manner. We note here that this
photon-atom bound state is present even when the atomic frequency
$\omega_{21}$ is placed outside the gap where the density of
states is not equal to zero and that is because of the special
singular behaviour around of the isotropic model there. For
``smoother" cases such as smoothed models for the PBG density of
modes or a Lorenzian density of modes both the corresponding
dressed states occur at complex frequencies which for long times
leads to complete decay of the population to the ground state.

\section{Conclusions}
In this introductory review article we have discussed phenomena in
photonic crystals with emphasis on the inhibition of spontaneous
emission of atoms in such materials. We started by describing the
physics behind photonic band gap formation by making a parallel
with the known phenomenon of electron localization in solids and
continued by presenting a simple model of a {\it band structure}
calculation. We  briefly presented the state of the art methods in
fabricating photonic band gap materials. Some of the exciting
applications in the field of optoelectronics were illustrated
followed by noting the basically theoretical but really exciting
possibility of {\it circuits of light} in all an optical computing
device. In the second part we moved to the quantum regime, i.e.,
the interaction of the quantized electromagnetic field inside a
photonic crystal with small atomic systems. We discussed the way
of treating the spontaneous emission of a two-level atom in free
space using the Weisskopf-Wigner theory. We concluded this
introductory review by presenting in detail the modification in
the spontaneous emission of the same two-level atom when embedded
in a photonic crystal of a specific type. New phenomena arising by
the localization of light as the complete {\it inhibition} of the
spontaneous decay, {\it atom-photon bound state} and {\it
population trapping} in a two-level atom were derived and
discussed.

 \section*{Acknowledgments}
D.G.A. acknowledges Prof. Artur Ekert  for useful discussions and
his support in the Cambridge CQC. E.P. and D.G.A would like to
thank Prof. Nikolaos Stefanou and Dr. Vassilios Yannopapas for
useful discussions in the area of photonic crystals. D.G.A. also
acknowledges St Catharine's College, University of Cambridge for
financial support and the Cambridge MIT Institute for travel
support. We acknowledge all the several groups that kindly allowed
their photos and graphs to be used in these article. We also
apologize if due to the introductory nature of this article, the
limited space and the vastness of the corresponding literature in
photonic crystals some groups with interesting contributions in
the field might have not been mentioned.

\section*{Biographies}

\subsection*{ Dr Dimitris G. Angelakis}
Dimitris G. Angelakis is a Research Fellow of St Catharine's
College, Cambridge position that he  holds at the Centre for
Quantum Computation in the Department of Applied Mathematics and
Theoretical Physics, University of Cambridge. He was awarded his
BSc degree and a MSc degree from the Physics Department,
University of Crete in 1997 and 1998, respectively and his PhD
from the Physics Department, Imperial College London in 2001. His
research interests include studies in the areas of quantum optics,
photonic crystals, CQED, Bell Inequalities tests, quantum
information theory and quantum computation.

\subsection*{ Dr Emmanouel Paspalakis}
Emmanuel Paspalakis received a BSc degree and a MSc degree from
the Physics Department, University of Crete in 1994 and 1996,
respectively, and a PhD degree from the Physics Department,
Imperial College London in 1999. After post-doctoral studies at
the Physics Department, Imperial College London he joined the
Materials Science Department, University of Patras in 2001, where
he is currently a Lecturer. His research contributions include
theoretical studies in the areas of quantum optics,
optoelectronics, dynamics of nanostructures and quantum
computation.

\subsection*{ Prof. Peter L. Knight}
Peter Knight is Head of the Physics Department at Imperial College
London and has worked in quantum optics for many years. He was
awarded his D Phil from Sussex University in 1972 following
undergraduate studies at the same institution. He subsequently
held research appointments at the University of Rochester and at
Royal Holloway before joining Imperial in 1979. He is the current
President of the Optical Society of America, and Editor of the
Journal of Modern Optics (and of Contemporary Physics).

\end{document}